\documentclass[prl,twocolumn,showpacs,amsmath,amssymb]{revtex4}
\usepackage{graphicx}
\usepackage{dcolumn}
\usepackage{epsfig}

\newcommand{\be}{\begin{equation}}
\newcommand{\ee}{\end{equation}}
\newcommand{\bea}{\begin{eqnarray}}
\newcommand{\eea}{\end{eqnarray}}
\newcommand{\nn}{\nonumber}

\begin{document}

\title{Unquenched complex Dirac spectra at nonzero chemical potential: 
two-colour QCD lattice data versus matrix model}
\author {Gernot Akemann$\,{}^\sharp$ and Elmar Bittner$\,{}^\flat$}

\affiliation{${}^{\sharp}$Department of Mathematical Sciences, Brunel 
          University West London, Uxbridge UB8 3PH, United Kingdom\\
       ${}^\flat$Institut f\"ur Theoretische Physik, Universit\"at Leipzig,
        Augustusplatz 10/11 D-04109 Leipzig, Germany}

\date   {\today}

\begin{abstract}
We compare analytic predictions of non-Hermitian chiral random matrix theory 
with the complex Dirac operator eigenvalue spectrum of two-colour lattice 
gauge theory with dynamical fermions at nonzero chemical potential.
The Dirac eigenvalues come in complex conjugate pairs, making 
the action of this theory real, and positive for 
our choice of two staggered flavours.
This enables us to use standard Monte-Carlo in testing the influence of 
chemical potential and quark mass on complex 
eigenvalues close to the origin. 
We find an excellent agreement between the analytic predictions 
and our data for two different volumes over a range of chemical potentials 
below the chiral phase transition. In particular we 
detect the effect of unquenching 
when going to very small quark masses. 
\end{abstract}



\maketitle

Non-Hermitian operators appear in many different areas of Physics:
in S-matrix theory with absorption \cite{FS}, in neural network and 
dissipative quantum dynamics \cite{GHS88}, 
disordered systems with imaginary vector potential \cite{Efetov}, 
Quantum Chromodynamics (QCD) with a  non-zero chemical potential 
$\mu$ \cite{Steph}, or $\theta$-vacuum term \cite{AGL}. 
Studying the influence of these non-Hermiticity parameters
may serve to a better understanding of fundamental problems in
nature, such as the deconfinement transition observed 
in heavy-ion collisions, or the strong CP problem.
In the present work we use an analytically solvable 
random matrix model (MM) as an effective theory
based on symmetries, and focus 
on the applications to QCD at $\mu\neq0$.
Given the wide range of complex 
MM applications \cite{FS} we expect our approach 
to be relevant to other problems in the same symmetry
class.

Lattice Gauge Theory (LGT) is one of the most prominent methods to study 
nonperturbative QCD. However, the introduction of a  
chemical potential
poses a difficult problem: the Dirac operator and thus the action becomes 
complex non-Hermitian. In general this invalidates standard Monte-Carlo 
techniques using importance sampling, the so-called sign problem. 
The quenched approximation, eliminating the complex fermion determinant, 
has been explained to fail using a MM approach \cite{Steph}. 
It maps QCD to a theory with conjugate quarks, with 
chiral symmetry unbroken in the massless limit 
(see \cite{OSV} for a recent discussion, including phase quenched MM).

Since the role of dynamical fermions is crucial  
our goal is to show that MM can predict unquenched lattice data at 
$\mu\neq0$. 
Although analytical MM results exist for unquenched QCD 
\cite{JAOSV}, a comparison to LGT remains an open challenge, due to  
a complex valued density with highly oscillatory 
regions \cite{JAOSV,OSV}. 
The various ways to avoid the sign problem including reweighting, Taylor
expansion, imaginary chemical potential or combinations of these
reviewed in \cite{murev} rely essentially 
on being close to the chiral phase transition at high temperature 
$T\approx T_c$. 
These techniques do not apply easily at the region of our interest at
small $T\approx0$ and $\mu$, where MM are conjectured to coincide with a
particular limit of the underlying effective chiral Lagrangian, the
$\epsilon$-regime ($\epsilon\chi$PT). 
MM are a different realisation of the group integral over constant
Goldstone modes at $\mu\neq0$ for QCD \cite{SV,AFV}, and they 
clearly distinguish   
the three different chiral symmetry breaking ($\chi$SB)
patterns at $\mu\neq0$
\cite{HOV}. Thus they may serve as a test for algorithms 
solving the sign problem.
In order to access in this region with dynamical fermions 
we follow the suggestion \cite{HKLM}: 
gauge group $SU(2)$ or the adjoint representation 
lead to a real action
as complex Dirac eigenvalues 
come in conjugated pairs. We have chosen $SU(2)$ 
with 2 staggered fermions 
because in this symmetry class MM predictions are available \cite{A05}.
The difficulty is that large masses $m$ effectively quench the Dirac 
eigenvalues close to the origin. Unquenching is seen only 
for small $m$ depending on the particular size of the system.
Our preliminary results for $SU(2)$ have been presented in
\cite{ABLMP}, compared to $SU(2)$ at $\mu\!=\!0$  \cite{BMW} and 
quenched QCD at $\mu\neq0$ \cite{AW,OW}.

We start by introducing the MM \cite{A05}, the complex extension 
of the chiral Gaussian Symplectic Ensemble, and its predictions. For an 
arbitrary number of $N_f$ quark flavours with masses $m_f$ it is defined as:
\bea
{\cal Z}_\nu(\{m_f\};\mu) &=&
\!\!\!\!\!\!\!\!\! \int\limits_{\mathbb{H}^{(N+\nu)\times
    N}}\!\!\!\!\!\!\!\!\! d\Phi  d\Psi 
\prod_{f=1}^{N_f} 
\det\!\left[ \begin{array}{cc}
m_f\mbox{\bf 1} & i \Phi + \mu \Psi \\
i \Phi^{\dagger} + \mu \Psi^{\dagger} & m_f\mbox{\bf 1}
\end{array} \right] \nonumber\\
&&\times\exp\left[-N\mbox{Tr} (\Phi^\dag \Phi\ +\ \Psi^\dagger\Psi)\right].
\label{Z2MM}
\eea
The matrices $\Phi$ and $\Psi$ of size $(N+\nu)\times N$ 
contain quaternion real matrix elements
without being self-dual,
\mbox{\bf 1} is the quaternion unity element. The Gaussian average of 
variance $\sigma\!=\!1/\sqrt{2}$
replaces the gauge average,
where we have fixed topology or
equivalently the number of exact zero eigenvalues 
$\nu\geq0$. The total number of eigenvalues $2N\!\sim\!V$ 
relates to the volume.
For $\Psi\!=$ \mbox{\bf 1} this MM was shown \cite{HOV} to be 
in the symmetry class of the adjoint representation, or gauge group
$SU(2)$ with staggered fermions, 
as in our case. 
In \cite{HOV}, the qualitative differences between the 3  $\chi$SB
classes were compared numerically.
Assuming universality, which is true for the MM of QCD \cite{SV,AFV,JAOSV},
we choose the $\mu$-part $\Psi\neq$ \mbox{\bf 1}, with
the same symmetry as the kinetic part.
This choice in eq.~(\ref{Z2MM}) permits an eigenvalue representation
with Jacobian 
$\sim \prod_{k>l}^N |z_k^2-z_l^2|^2|z_k^2-z_l^{\ast\,2}|^2
\prod_{h=1}^N |z_h^2-z_h^{\ast\,2}|^2$
\cite{A05},
and mass term $\prod_{f=1}^{N_f} 
|z_j^2+m_f^2|^2$ (here $z\!=\!x+iy$ is real at $\mu\!=\!0$). 
Due to the Jacobian complex eigenvalues are repelled 
both from the $x$- and $y$-axis, the signature of this $\chi$SB class.
The QCD MM \cite{JAOSV} has no such repulsion.

In \cite{A05} all complex eigenvalue correlations were derived in a
closed form as the Pfaffian of a matrix kernel of skew orthogonal
polynomials on $\mathbb{C}$. We only give 
the density for $N_f\!=\!2$ of equal mass needed here. 
For a discussion of matching MM and staggered flavours we refer to the first
of Ref. \cite{BMW}. No $2nd$ or $4th$ root is taken here. 
Taking the large-$N$ limit at weak non-Hermiticity \cite{FS}
the eigenvalues, masses {\it and} $\mu$ are rescaled as
\be
z\cdot2N\sigma \equiv\xi, \ m\cdot2N\sigma \equiv\eta,\ \ 
\lim_{N\to\infty,\ \mu\to0}\mu^2\cdot 2N\equiv\alpha^2 \ ,
\label{weak}
\ee
resulting in the usual MM
level spacing $\pi$. 
The $\epsilon$-regime of the corresponding chiral Lagrangian
\cite{Kogut} dictates the same scaling with two {\it different} constants: 
$V\Sigma m$ and $F_\pi^2V\mu^2$, 
the chiral condensate $\Sigma$ and decay
constant $F_\pi$, respectively. 
This implies the existence of one free fit parameter in 
$\alpha^2$ in our model.
This observation has very recently been proposed to measure $F_\pi$ 
for QCD using isospin chemical potential \cite{poulF}. 
We note that MM predictions at $\mu\!=\!0$ are parameter-free in
units of $\Sigma$.

In terms of the rescaled variables  eq.~(\ref{weak}) 
the spectral density defined as 
$\rho(z)\equiv\langle\prod_{j=1}^N\delta^{(2)}(z-z_j)\rangle$ 
is given by 
\be
\rho^{(N_f=0)}(\xi) =
(\xi^{\ast\,2}-\xi^2)w(\xi,\xi^*)\kappa(\xi,\xi^*)
\label{rhoquench}
\ee 
for quenched. The rescaled weight and kernel read
\bea
w(\xi,\xi^*)&=& \frac{1}{32\alpha^4}|\xi|^{2} 
K_{2\nu}\left(\frac{|\xi|^2}{2\alpha^2}\right)
\mbox{e}^{\frac{1}{4\alpha^2}(\xi^2+\xi^{*\,2})},
\label{weight}\\
\kappa(\xi,\zeta^*)&=&\int_0^1\! ds\! \int_0^1\! 
\frac{dt}{\sqrt{t}}\ \mbox{e}^{-2s(1+t)\alpha^2}
\nn\\
&\times&
(J_{2\nu}(2\sqrt{st}\xi)J_{2\nu}(2\sqrt{s}\zeta^\ast)
-
(\xi\leftrightarrow\zeta^*)
),
\eea
\begin{figure}[h]
\centerline{
  \epsfig{figure=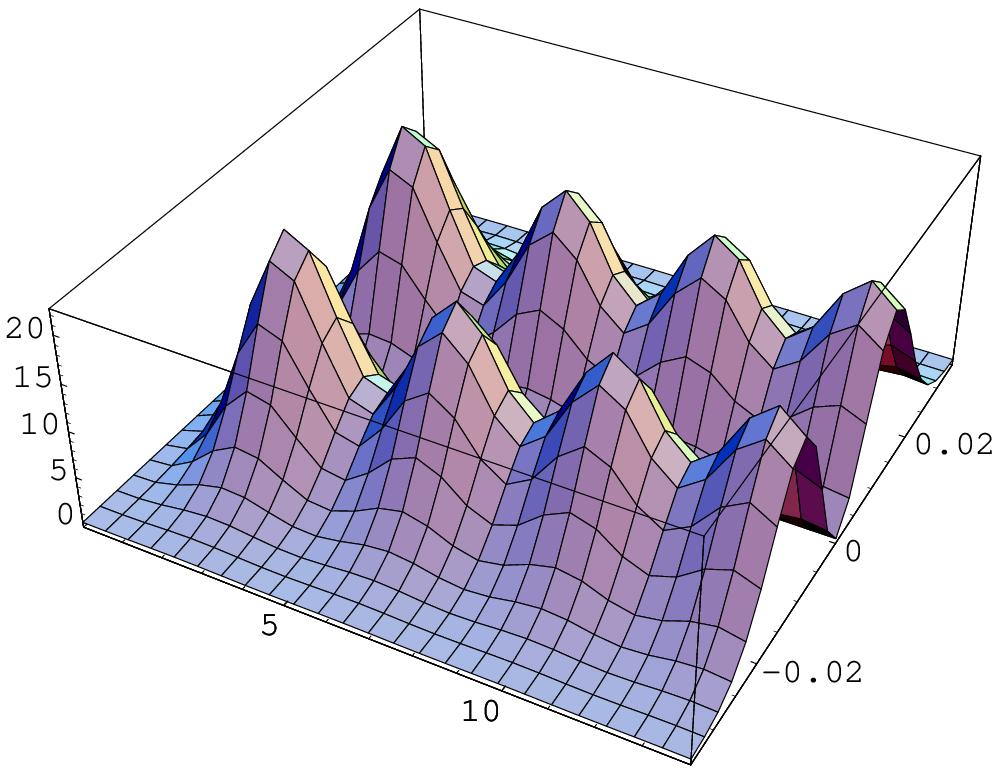,height=33mm}\hspace*{-1mm}
  \epsfig{figure=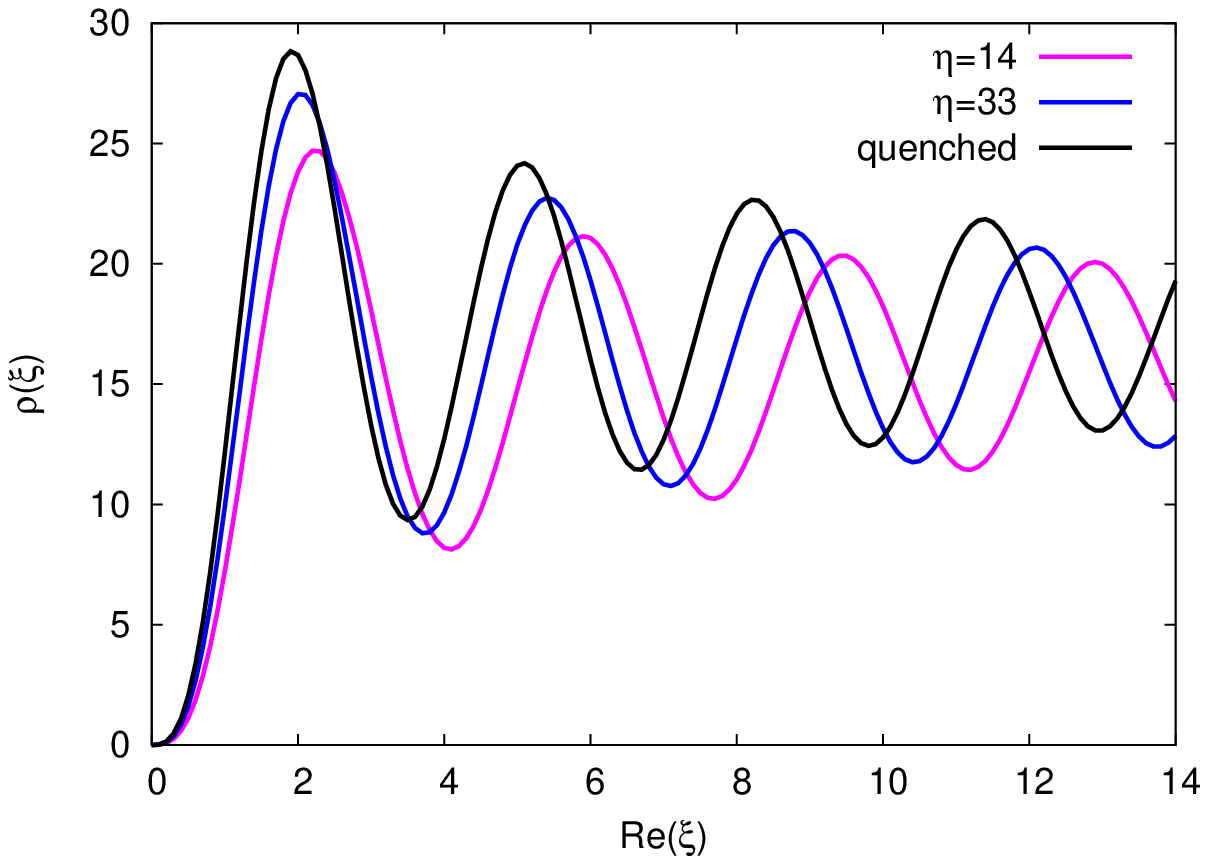,height=33mm}
 \put(-240,10){\tiny ${\rm Re}(\xi)$}
 \put(-150,25){\tiny ${\rm Im}(\xi)$}
 \put(-260,65){\tiny $\rho(\xi)$}
}
\caption{The spectral density with $N_f\!=\!2$ eq.~(\ref{rhodyn}) at 
$\alpha\!=\!0.012$, $\eta\!=\!8.74$ and $\nu\!=\!0$ (left), 
and a cut parallel to the $x$-axis 
for different masses (right).\\[-7ex] 
}
\label{3dweak}
\end{figure}
\noindent
containing $K$- and $J$-Bessel functions.
The weight for the eigenvalues is non-Gaussian due to the 
non-trivial decoupling of eigenvectors.
The unquenched density
\bea
\rho^{(N_f=2)}(\xi) &=&
\rho^{(0)}(\xi)-
(\xi^{\ast\,2}-\xi^2)w(\xi,\xi^*)\nn\\
&&\times
\frac{|\kappa(\xi,i\eta)|^2-|\kappa(\xi^*,i\eta)|^2}{\kappa(i\eta,i\eta)},
\label{rhodyn}
\eea
with a mass dependent correction term is shown in Fig.~\ref{3dweak} left
(for more details, see \cite{A05}).
Compared to $\mu\!=\!0$ in \cite{BMW}
the peaks locating the eigenvalues are split in two, the particularity 
of this $\chi$SB class.
A cut parallel to the $x$-axis along the maxima illustrates 
the effect of dynamical flavours at increasing $\eta$. 
At value $\eta\approx100$ the smallest eigenvalues are effectively quenched,
following eq.~(\ref{rhoquench}).
\begin{figure}[h]
\centerline{
  	\epsfig{figure=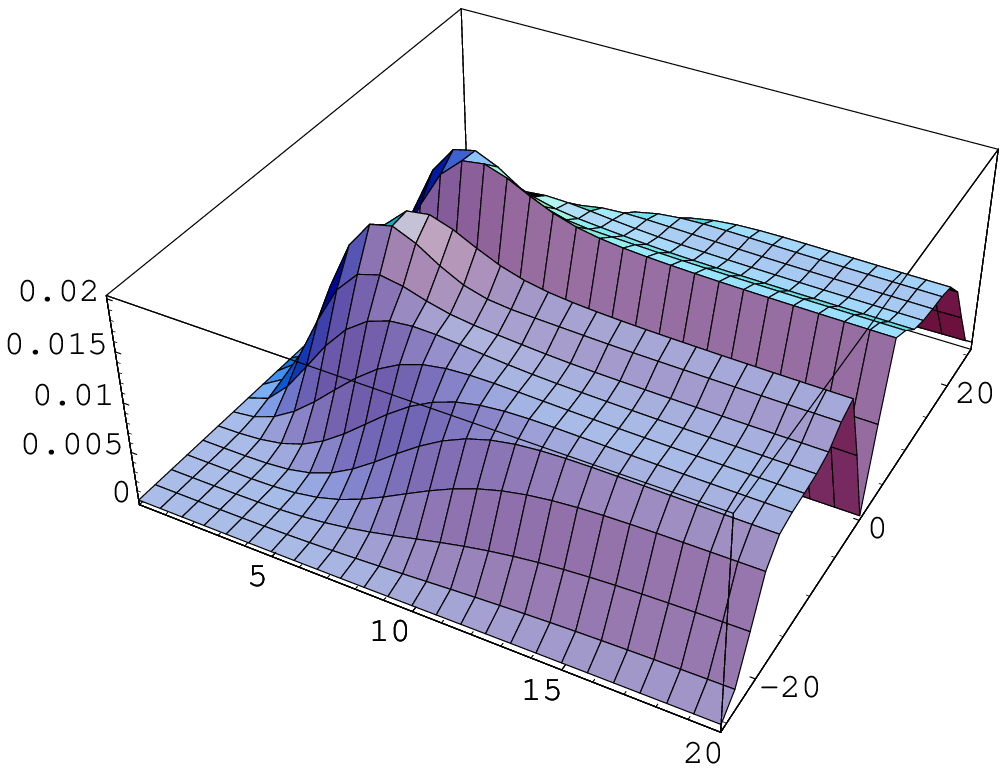,height=33mm}\hspace*{-1mm}
	\epsfig{figure=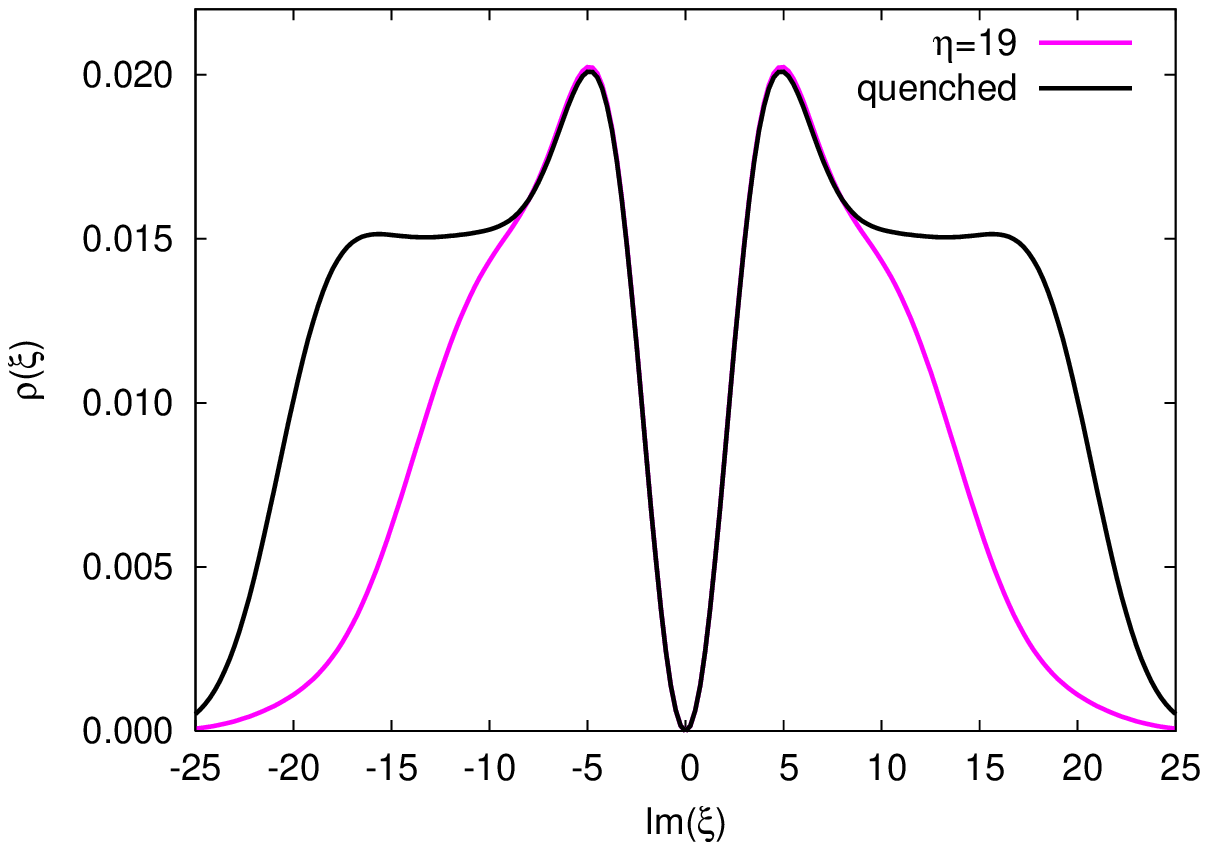,height=33mm}
 \put(-230,10){\tiny ${\rm Re}(\xi)$}
 \put(-145,30){\tiny ${\rm Im}(\xi)$}
 \put(-255,70){\tiny $\rho(\xi)$}
}
\caption{
The density eq.~(\ref{rhodyn}) 
at $\alpha\!=\!3.28$, $\eta\!=\!19$, and $\nu\!=\!0$ (left), and a cut
parallel to the $y$-axis (right) vs. quenched.\\[-4ex] 
}
\label{3dstr}
\end{figure}
Increasing $\alpha$ rapidly damps the oscillations. In Fig.~\ref{3dstr} 
the density is now constant along a stripe, as expected from mean
field. The effect of the additional zeros of $\rho^{(N_f=2)}(\xi)$ at 
$\xi\!=\!\pm i\eta$ in the left
corners of Fig.~\ref{3dstr} are clearly visible. 
Here we choose a cut in 
$y$-direction on the maxima to highlight the difference to quenched.
At this value of $\alpha$  the perpendicular cut in $x$-direction 
shows no dependence on $\eta$, in contrast to the previous figure.
In the strong non-Hermiticity limit $\alpha\to\infty$
the plateau extends to $\infty$ in both $x$- and $y$-direction 
\cite{A05}. 
Away from the edge Fig.~\ref{3dstr} is already very close to
$\alpha\!=\!\infty$, to which we compared our preliminary data 
\cite{ABLMP}. 

\begin{table*}[t]
  \vspace*{-3mm}
  \caption{Summary of simulation parameters at $\beta=1.3$}
  \label{param}
  \begin{ruledtabular}
    \begin{tabular}{lllllllll}
      $V$   & $\mu$    					& mass $ma$ 	&level
      spacing $d$ 			& $\eta$  	& $\alpha$ &
      $F_\pi^2/\Sigma$ 
      & $E_{Th}$	 	& no. of config.\\[1mm]  
      $6^4$ & $1.0\cdot 10^{-3}$ 	& 	$0.035$		& $8.0(2)
      \cdot 10^{-3}$ 	& $13.7$ 	& $0.0124(3)$	& $0.39(3)$
		&	$\sim 17$ & $45\,000$ \\ 
      $6^4$ & $1.0\cdot 10^{-3}$ 	& 	$0.07$		& $6.6(2)
      \cdot 10^{-3}$ 	& $33.3$		& $0.0127(3)$	& $0.34(3)$
		&	$\sim 25$	& $15\,000$ \\ 
      $6^4$ & $1.0\cdot 10^{-3}$ 	& 	$20$			&
      $5.0(2) \cdot 10^{-3}$ 	& $12\,566$	& $0.0153(3)$ & $0.37(3)$
		&	$> 100$		& $10\,000$ \\ 
      $6^4$ & $0.2$               	& 	$0.06$		& $9.9(1)
      \cdot 10^{-3}$ 	& $19.0$		& $2.21(4)$		&
      $0.38(1)$ 
      &$\sim 10$	& $20\,000$ \\ 
      $6^4$ & $0.2$               	& 	$20$			&
      $4.5(1) \cdot 10^{-3}$ 	& $13\,962$	& $3.28(4)$	& $0.38(2)$
		& $> 200$		& $20\,000$ \\
      $8^4$ & $5.625\cdot 10^{-4}$	& 	$0.035$		& $3.0(3)
      \cdot 10^{-3}$ 	& $35.5$		& $0.0107(9)$	& $0.35(5)$
 		& $\sim 20$ & $8\,000 $ \\ 
      $8^4$ & $5.625\cdot 10^{-4}$	& 	$0.07$		& $2.1(1)
      \cdot 10^{-3}$ 	& $104.7$	& $0.0124(3)$	& $0.32(4)$
		& $\sim 35$ & $5\,000 $ \\ 
      $8^4$ & $5.625\cdot 10^{-4}$	& 	$20$			&
      $1.6(1) \cdot 10^{-3}$ 	& $39\,270$	& $0.0152(3)$	& $0.37(4)$
		& $> 35$ & $3\,000 $ \\ 
      $8^4$ & $0.1125$             	& 	$20$			&
      $1.4(1) \cdot 10^{-3}$ 	& $44\,880$	& $3.29(4)$ & $0.38(5)$
		& $> 40$			& $6\,000 $ \\
    \end{tabular}
  \end{ruledtabular}
  \vspace*{-3mm}
\end{table*}
We now turn to the LGT side, describing the details of our simulations
summarised in 
Table \ref{param}.
Our data were generated for gauge group $SU(2)$ 
with $N_f\!=\!2$ unimproved staggered fermion flavours of equal mass at 
coupling constant $\beta\!=\!4/g^2\!=\!1.3$, using the code of 
\cite{HKLM}. In this setup the fermion
determinant remains real and standard Monte Carlo applies.
Our choice of parameters was dictated by several
constraints:
i) 
Wilson fermions explicitly break chiral symmetry and are already complex 
at $\mu\!=\!0$, making the effect of $\mu\neq0$ difficult to 
disentangle. At $\mu\!=\!0$ Ginsparg-Wilson type fermions have been shown to
exactly preserve chiral symmetry and topology (see \cite{EHKN} for references
and a comparison to MM).
Similar results have been obtained with 
improved staggered fermions for QCD \cite{EF}. 
Apart from being very expensive it is
conceptually not clear so far how to extend these results to $\mu\neq0$.
Thus we use the Dirac operator of 
standard, unimproved staggered fermions: 
\begin{eqnarray}
  \label{Dirac}
&&  D_{x,y}=
  \frac{1}{2a}
\Big[
 \sum\limits_{\gamma=\hat{x},\hat{y},\hat{z}}
 \left(U_{\gamma}(x)\kappa_{\gamma}(x)\delta_{y,x\!+\!\gamma}-{\rm h.c.}\right)
\\
  &&
 +\!\!\left(U_{\hat{t}}(x)\kappa_{\hat{t}}(x)e^{\mu}
    \delta_{y,x\!+\!\hat{t}}\!
    -\!U_{\hat{t}}^{\dagger}(y)\kappa_{\hat{t}}(y)
    e^{-\mu}\delta_{y,x\!-\!\hat{t}}\right)\!\!\Big],
\nonumber
\end{eqnarray}
with link variables $U$ and staggered phases $\kappa$.
Consequently our simulations are topology blind, and we set $\nu\!=\!0$ in the
following. We note that even at $\mu\!=\!0$  
the transition from staggered to the correct continuum symmetry class has not 
yet been observed.
ii) 
A proper resolution of complex eigenvalues requires
of the order of $10^4$ configurations for each parameter set. Even
with cheap staggered fermions we are confined to small lattices
$V\!=6^4$ and $8^4$.
iii) 
We want to make the window where MM apply as large as
possible. At the Thouless energy scale \cite{OV}, 
$E_{Th}\sim F_\pi^2/\Sigma\sqrt{V}$, 
the Dirac eigenvalues start to feel the propagating Goldstone
modes, and leave the $\epsilon$-regime.
Increasing $\beta$ (decreasing $F_\pi$) 
at fixed $V$ fewer Dirac eigenvalues follow MM
statistics. This leads us to rather 
strong coupling. 
iv) 
Small masses $m$ considerably
slow down the computation. 
Because of rescaling $\eta\!=\!m\Sigma V$
this effects our choice of the volume too.

To compare our data with the prediction eq.~(\ref{rhodyn})
we have to determine two parameters from the data.
First, we measure the mean 
level-spacing $d\sim1/\rho(0)$, by using the Banks-Casher relation
at $\mu\!=\!0$, $\pi\rho(0)\!=\!\Sigma V$, where $\rho(0)$ is the mean spectral
density. We project the complex eigenvalues onto the real
axis and measure their distance.
Due to $\mu\ll1$ the 2d geometric distance between eigenvalues
agrees within errors.
This provides the volume rescaling
of lattice eigenvalues, masses and $\mu^2$:
\begin{equation}
za\cdot\pi /d \ \equiv \xi\ , \ \ 
ma\cdot\pi/d \ \equiv \eta \ , \ \ 
\mbox{and} \ \ \mu^2C\cdot\pi/d\equiv \alpha^2 .
\label{rescale}
\end{equation}
Second, the remaining constant $C\sim
F_\pi^2/\Sigma$ 
is obtained by a fit to the data.
We cut the data rescaled as in eq.~(\ref{rescale}) parallel to the
$y$-axis on the first maxima, which are easy to identify for all $\alpha$. 
Then we fit to the integral of the analytic curve,
thus eliminating the choice of
bins in this direction (for clarity we show 
histograms in all Figures). This procedure avoids cumbersome 3d fits and 
fixes the normalisation of the first maxima.
\begin{figure}[h]
\centerline{
    \epsfig{figure=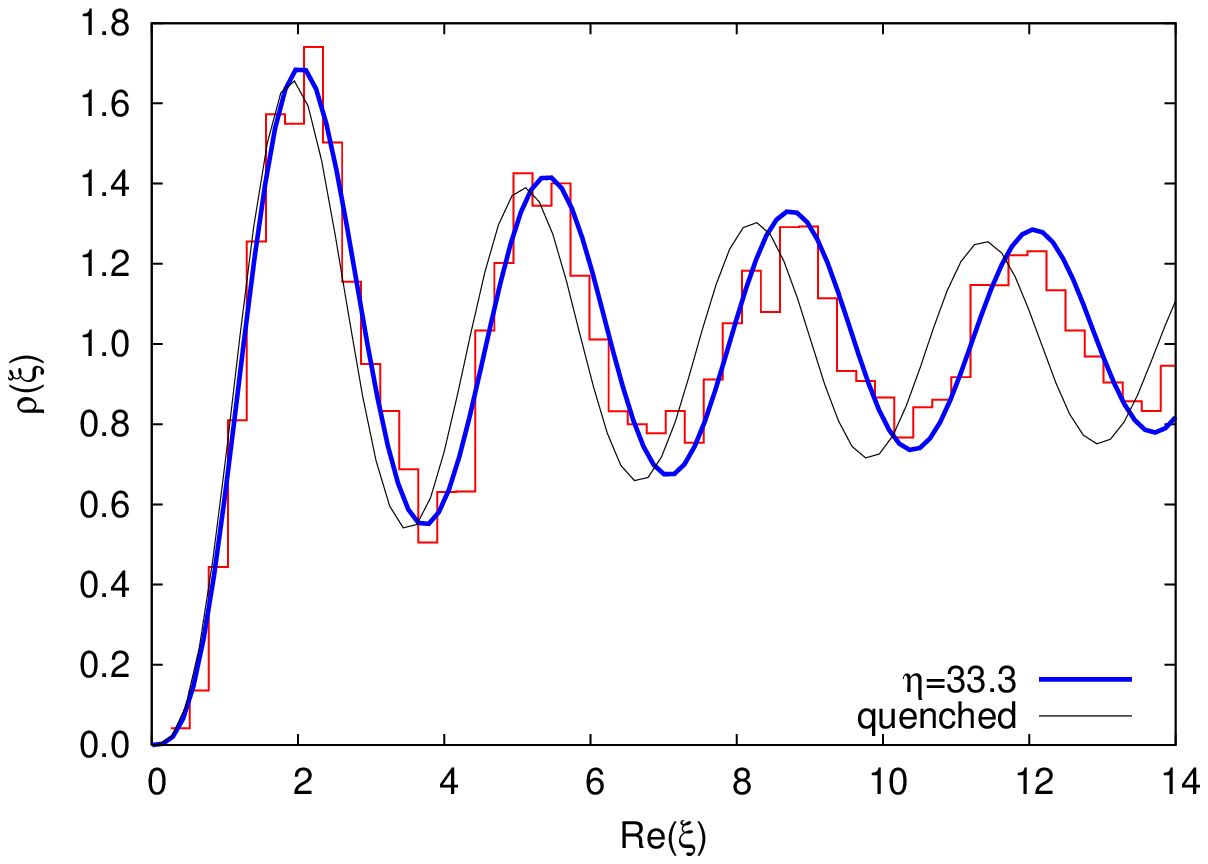,height=33mm}\hspace*{-1mm}
    \epsfig{figure=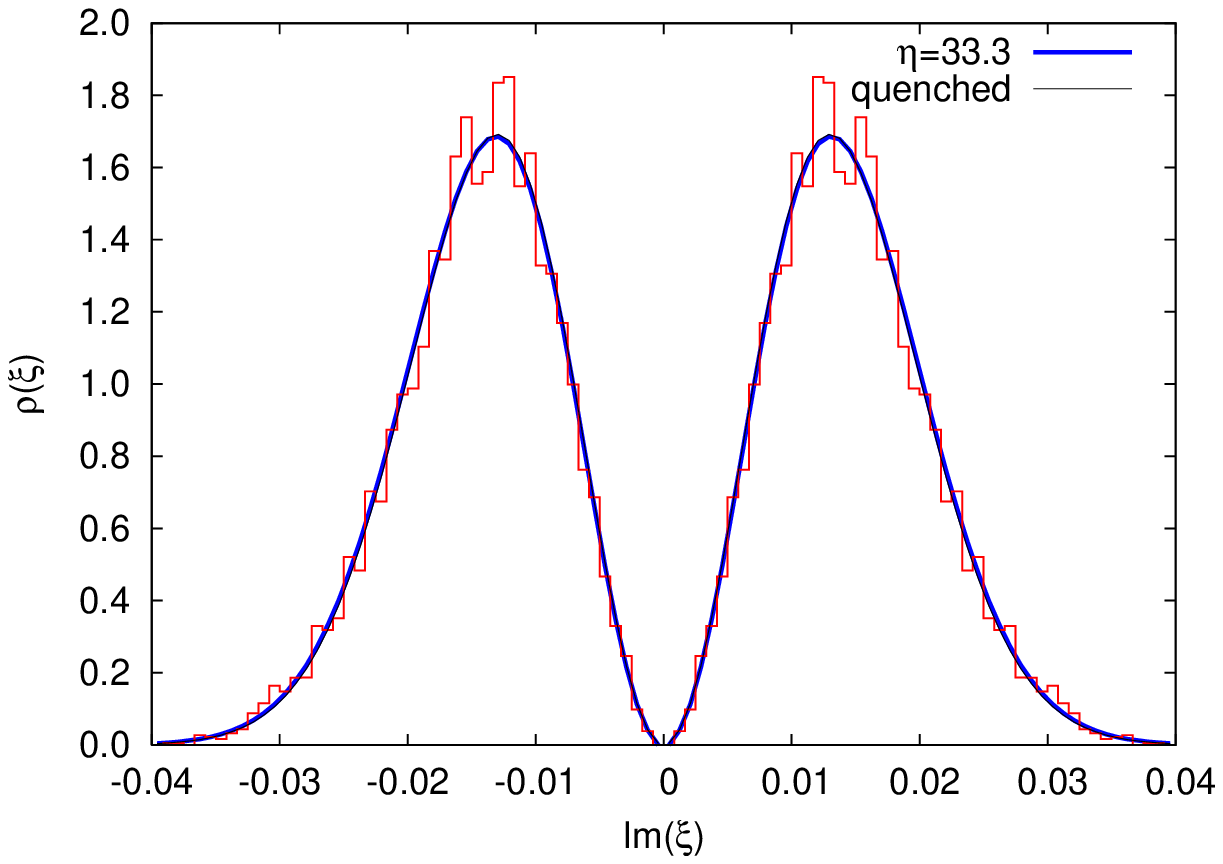,height=33mm} 
}
\centerline{
  \epsfig{figure=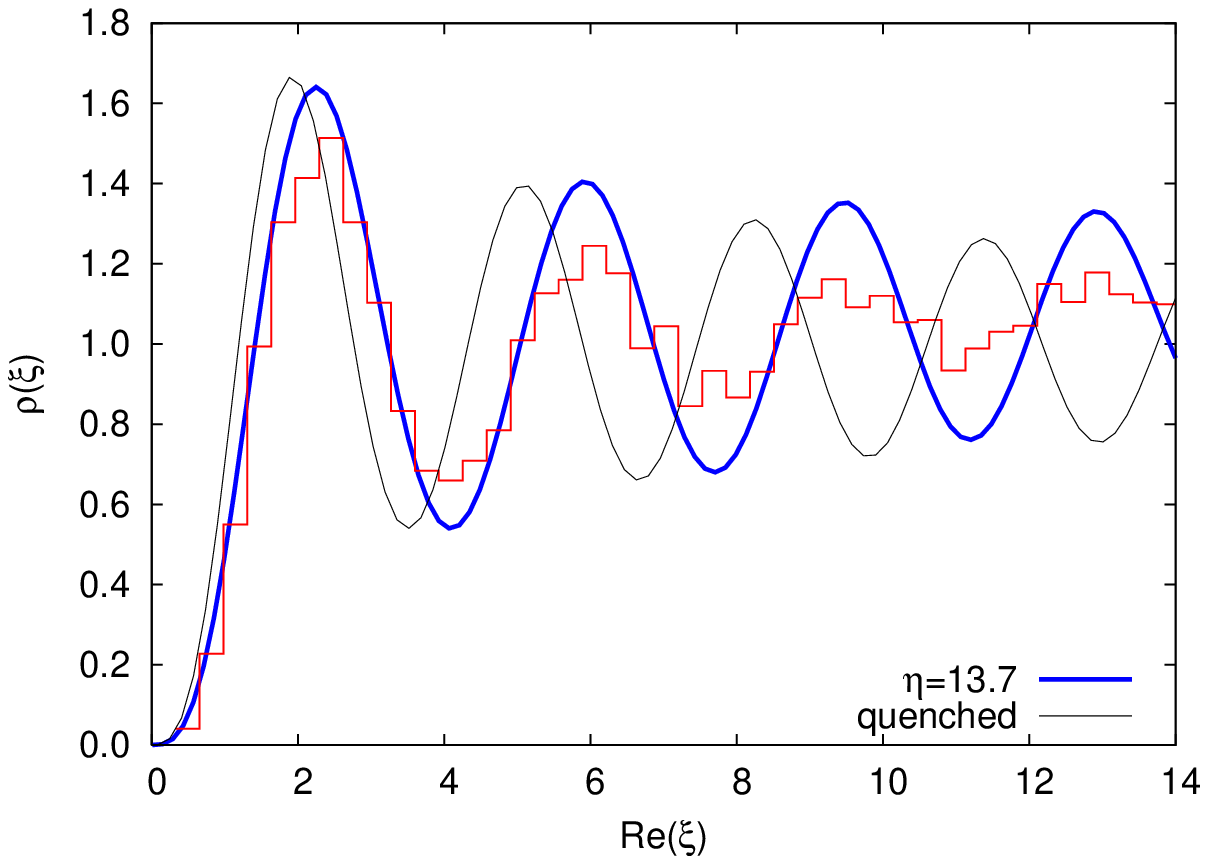,height=33mm}\hspace*{-1mm}
  \epsfig{figure=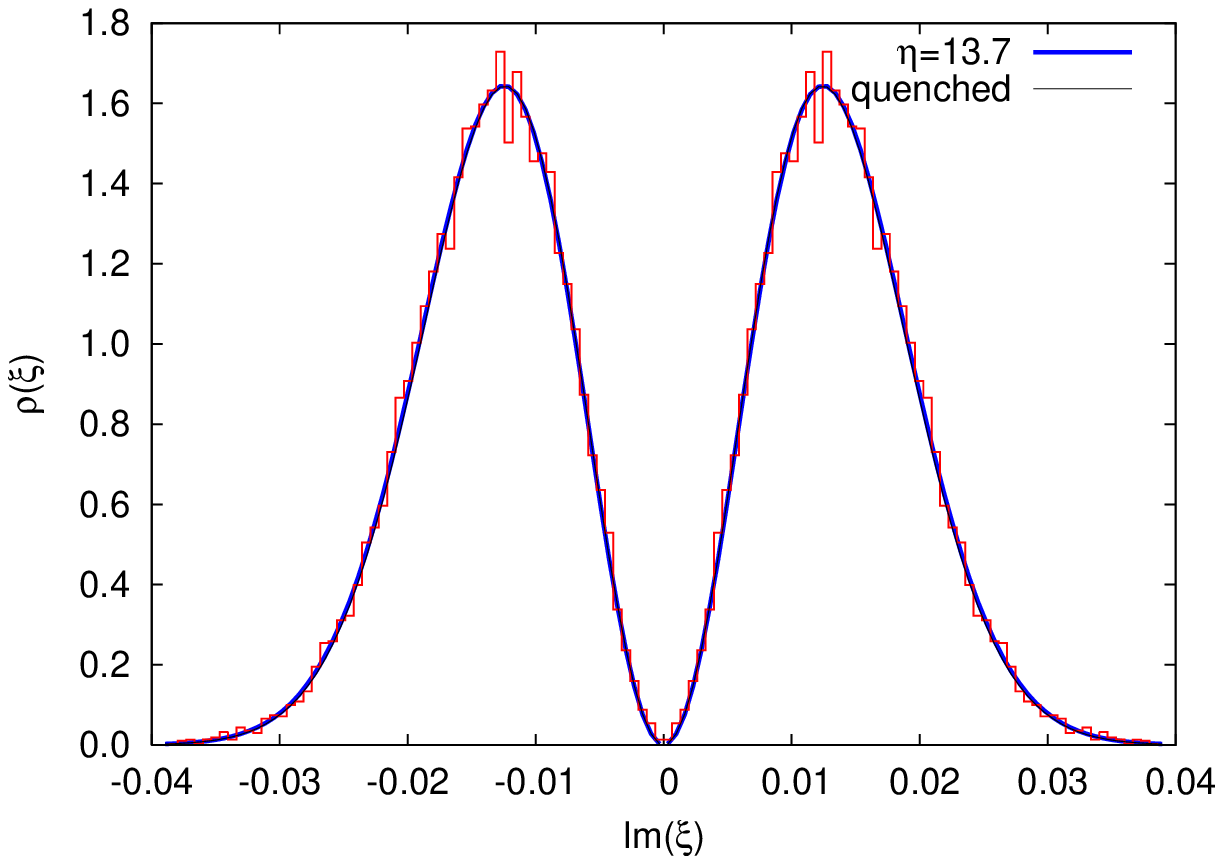,height=33mm}
}
\caption{
The effect of dynamical fermions at $\mu\!=\!10^{-3}$, $V\!=\!6^4$:
$\alpha\!=\!0.0127$, $\eta\!=\!33.3$ (upper), and 
$\alpha\!=\!0.0124$, $\eta\!=\!13.7$ (lower).\\[-7ex] 
}
\label{dyn1}
\end{figure}

We start with $\alpha\ll1$.
For mass $\eta\!=\!33.3$ in Fig.~\ref{dyn1} (up)
the data show an excellent match to the
prediction eq.~(\ref{rhodyn}) including the 4$th$ maximum, clearly deviating
from the quenched curve shown for comparison. 
The perpendicular cut on the first maxima (up right) 
is perfectly agreeing, allowing to fit $\alpha$ 
to an relative error of $\approx 2\%$. The 
repulsion from the $x-$axis correctly identifies this $\chi$SB class 
(QCD has a single peak, see \cite{AW}), and we have observed this pattern
down to $\mu\!=\!10^{-6}$. 
Note that fitting
$\alpha$ does not move the maxima in the left curves.
A similar data set for $V\!=\!8^4$ leads to the same conclusions 
(see Table 1).  
\begin{figure}[b]
 \centerline{ \epsfig{figure=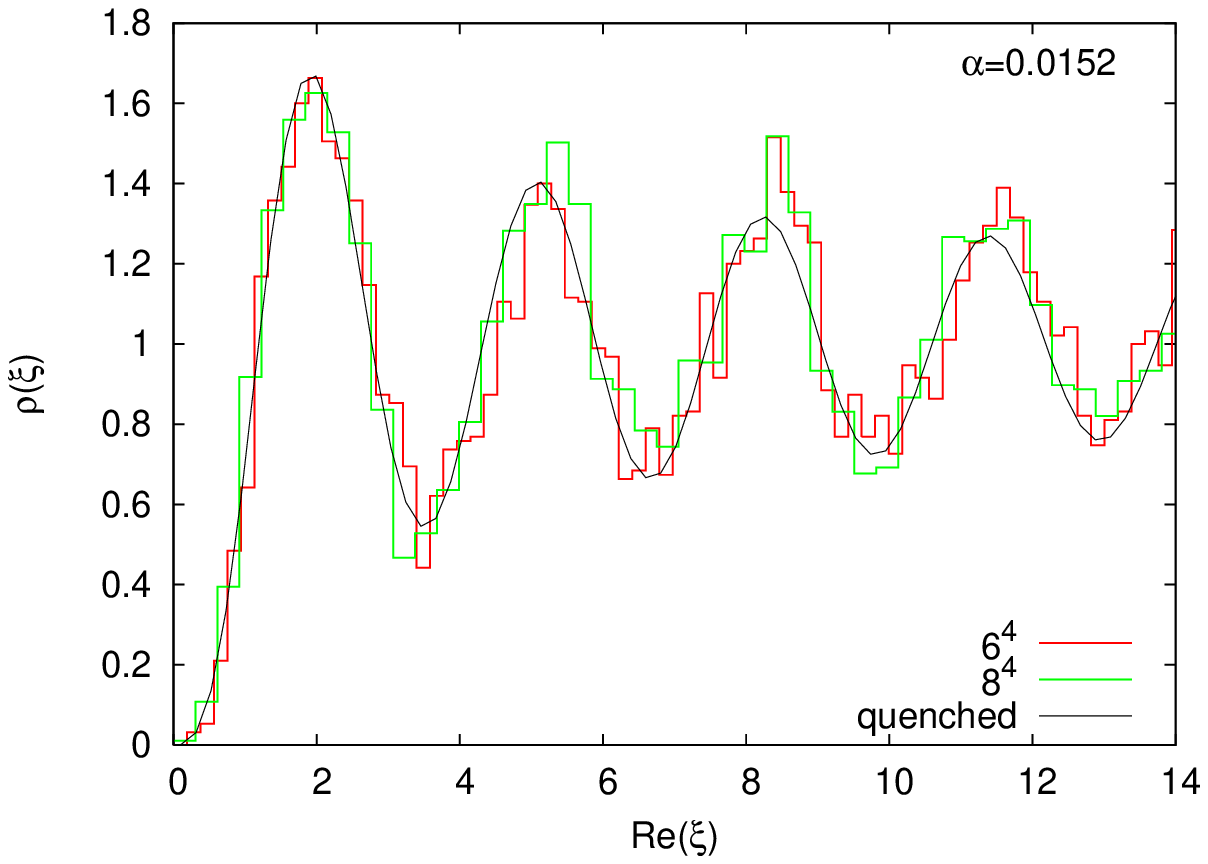,height=33mm}\hspace*{-1mm}
    \epsfig{figure=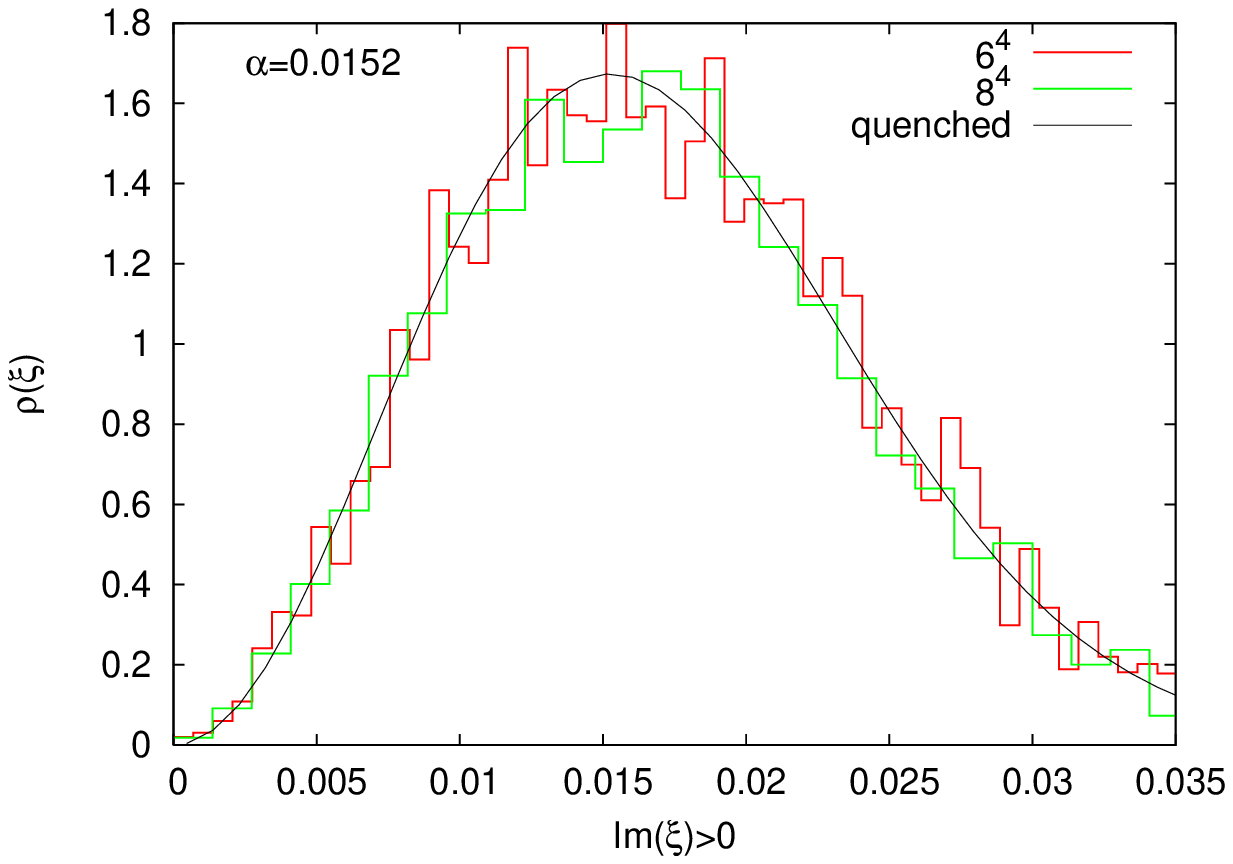,height=33mm} }
\caption{
The scaling $\mu_i^2V_i\!=$const.: 
$V_1\!=\!6^4$, $\mu_1\!=\!10^{-3}$ (green) vs. 
$V_2\!=\!8^4$, $\mu_2\!=\!5,625\cdot10^{-4}$ (red), both at $ma\!=\!20$  for
$y>0$.  
}
\vspace*{-3mm}
\label{scaling}
\end{figure}
Fig.~\ref{dyn1} (down) shows the smallest
mass we could simulate, with $\eta\!=\!13.7$. 
The data
follow the analytic curve including the 2$nd$ minimum, 
deviating substantially from quenched. 
The onset of $E_{Th}\!\sim\!17$ 
is responsible for a smaller window of agreement ($E_{Th}\!\sim\!25$ at 
$\eta\!=\!33.3$).
The perpendicular cut shows again perfect
agreement. 
Despite excessive statistics the
fluctuations on the maxima are still visible. 
To exclude that the observed deviations 
from quenched are finite volume effects we compare
to \cite{BMW} at $\mu\!=\!0$ and otherwise identical parameters. 
There finite volume shifts the data to the 
{\it left} of the MM curve. 
Thus for larger volumes
the deviation from quenched would be even more pronounced.
\begin{figure}[t]
\centerline{
  \epsfig{figure=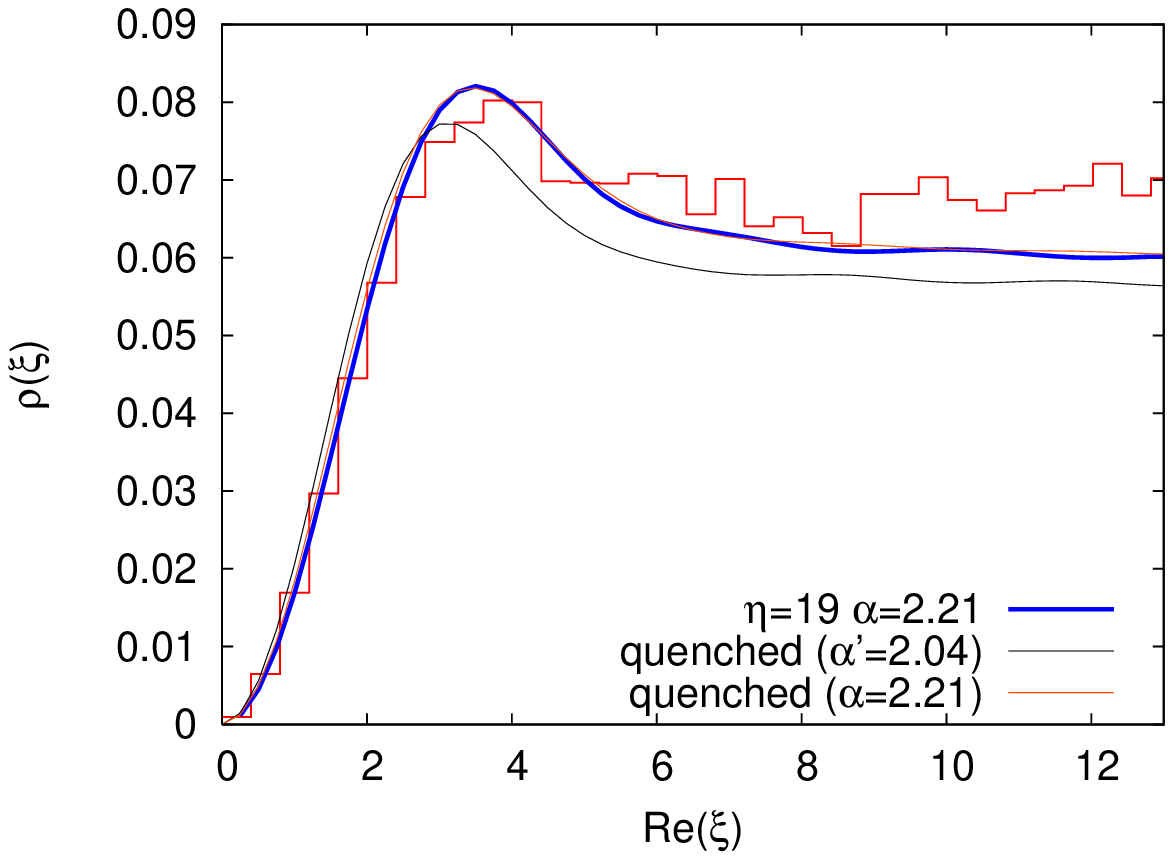,height=33mm}\hspace*{-1mm}
  \epsfig{figure=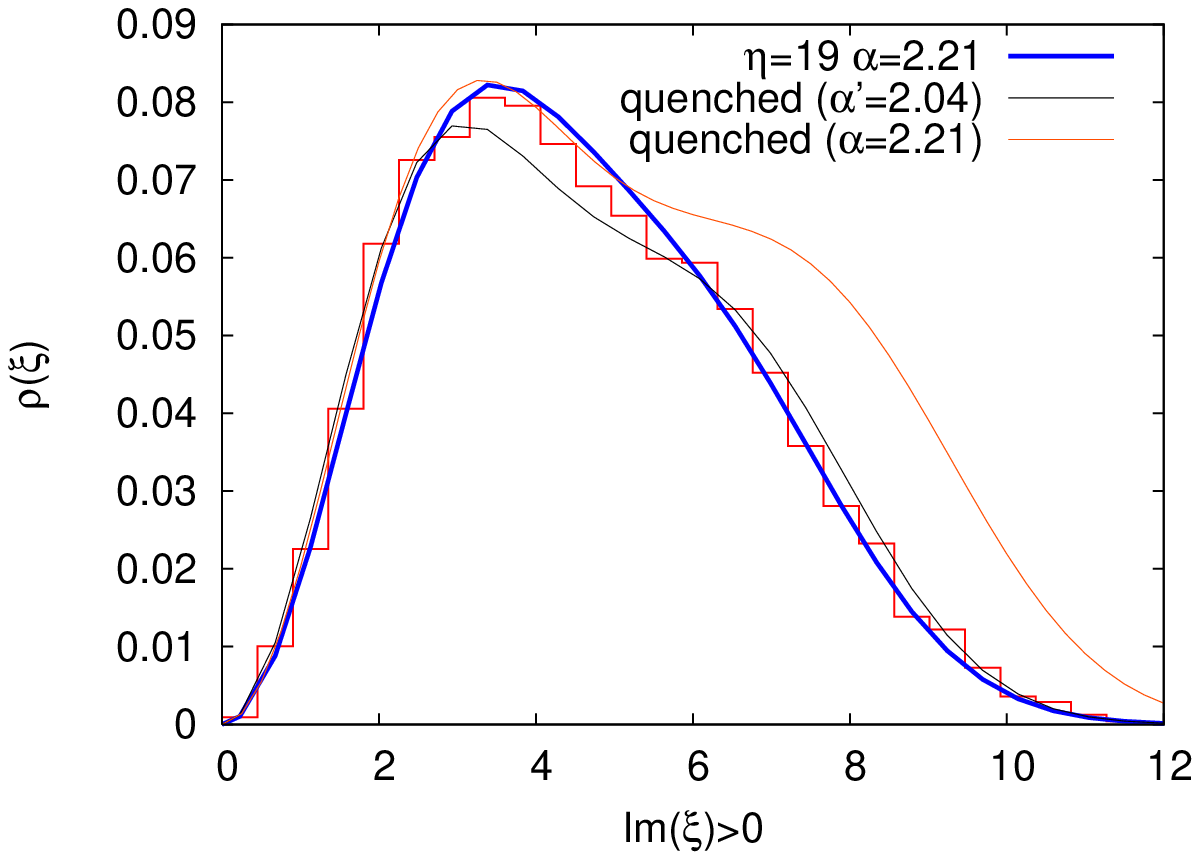,height=33mm}
}
\caption{
The effect of dynamical fermions at $\mu\!=\!0.2$, $V\!=\!6^4$: 
$\alpha\!=\!2.21$, $\eta\!=\!19.0$ for $y>0$. Note that
$E_{Th}\!\sim\!10$.\\[-7ex]  
}
\label{dyn2}
\end{figure}
Second we check the scaling hypothesis from MM and $\epsilon\chi$PT, 
choosing $\mu_1^2/\mu_2^2\!=\!6^4/8^4$ with $\mu_i^2V_i\!=$ const. 
For simplicity \cite{footnote1} 
we have chosen a large $\eta$ to effectively quench the small eigenvalues. In
Fig.~\ref{scaling} both sets of
data show an excellent agreement to the same curve from
eq.~(\ref{rhoquench}). 
Independent fits for $\alpha$ agree within errorbars. Here only 
$y\!>\!0$ is shown for a better resolution, the second peak at $y\!<\!0$
follows from symmetry.
The same results are obtained for smaller, unquenched masses (see Table 1).
 
\begin{figure}[t]
 \centerline{ \epsfig{figure=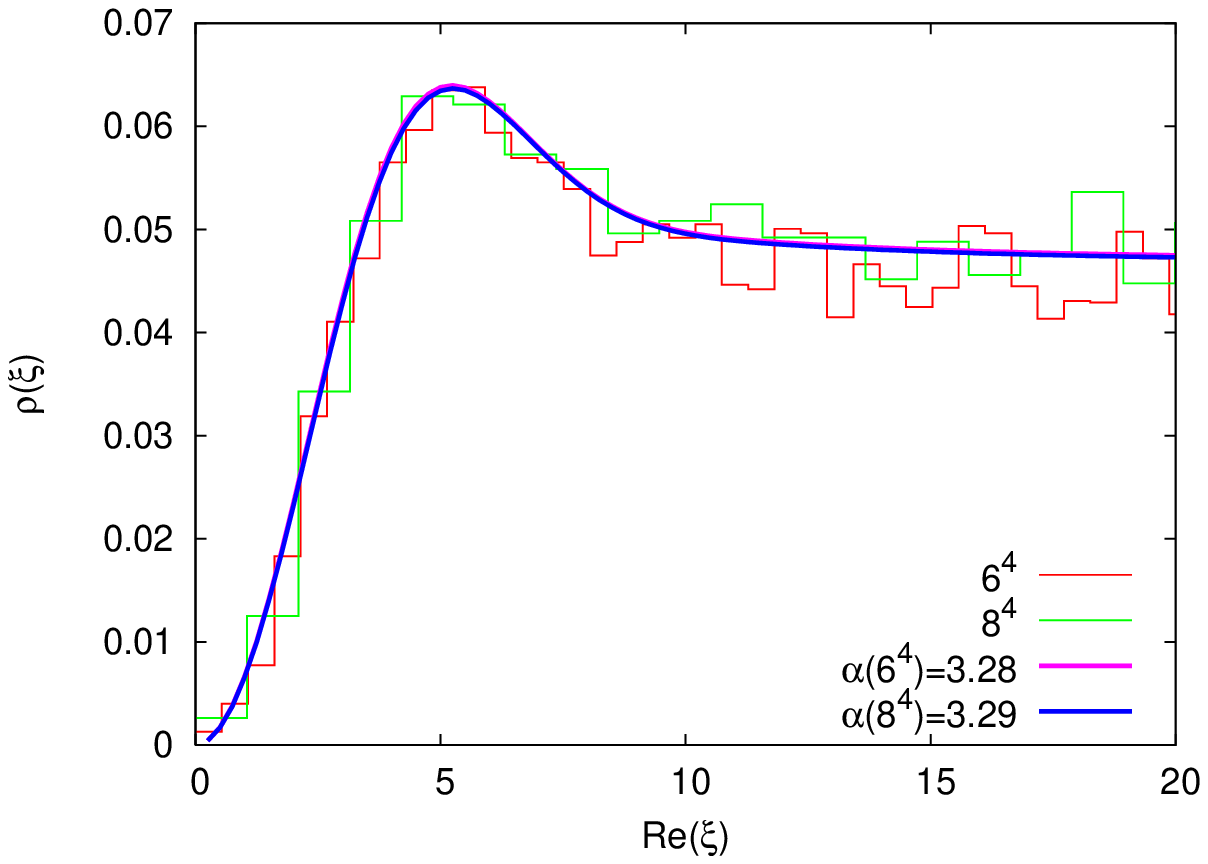,height=33mm}\hspace*{-1mm}
    \epsfig{figure=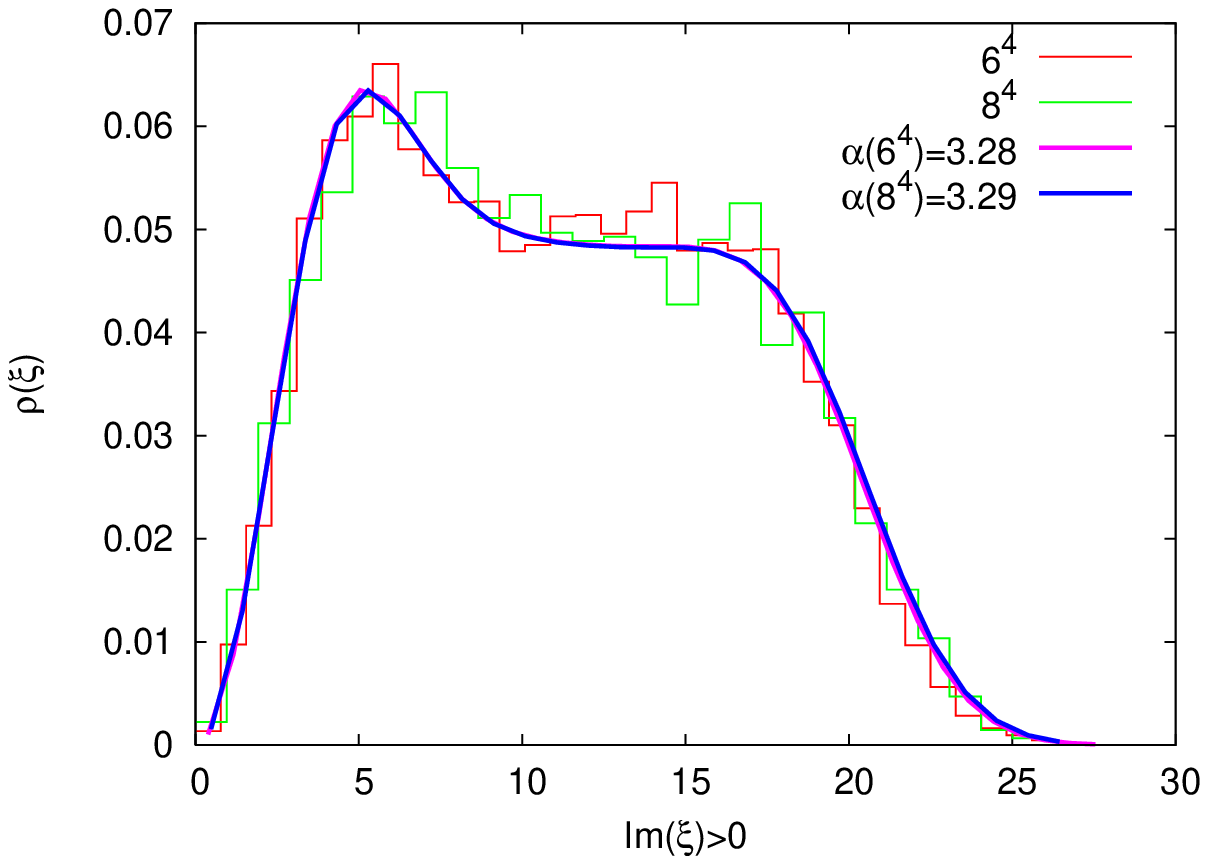,height=33mm} }
\caption{
The scaling of $\mu_i^2V_i\!=$const.: 
$V_1\!=\!6^4$, $\mu_1\!=\!0.2$ (red) vs. 
$V_2\!=\!8^4$, $\mu_2\!=\!0.1125$ (green), both at
$ma\!=\!20$ for $y>0$.\\[-6ex] 
}
\vspace*{-3mm}
\label{scaling2}
\end{figure}

Next we turn to a larger $\mu$, remaining below the
phase transition $\mu_c\approx 0.3$ found in \cite{BLMP} for $V\!=\!6^4$,
$\beta\!=\!1.3$. Above the transition at $\mu\!=\!0.4$ 
the data deviate from the MM as
expected. We first determine $d$, which fixes the scale and $\eta$.
At these parameter values unquenching is seen 
in $y$-direction (Fig.~\ref{dyn2} right), which is used to fit $\alpha$. 
Here both $\alpha$ and the given $\eta$ influence the width. 
We therefore compare the unquenched curve to  
the quenched one with $\alpha'\neq\alpha$ fitted independently (lower curve).
For comparison we also give 
the quenched curve at the same $\alpha$ (upper curve).
Again we display $y>0$ only. Clearly the unquenched curve 
describes the data best both in $x$- and $y$-direction.
The onset of $E_{Th}$ 
is responsible for the deviation between data and MM for $x>10$ in
Fig.~\ref{dyn2} (left). 
In Fig.~\ref{scaling2} 
the scaling of constant $\mu_i^2V_i$ is tested for these $\mu$-values,
where we
have again chosen $\eta_i\gg100$ to effectively quench.
Both data sets nicely follow the same curve, the fits for $\alpha$
agree within errorbars.

To conclude we have shown that our unquenched 
$SU(2)$ lattice data are quantitatively 
very well described by complex MM predictions over a wide range of chemical
potentials and masses.
We find agreement with $\epsilon\chi$PT by confirming 
the scaling $F_\pi^2V\mu^2$, with a consistent value
for $F_\pi^2/\Sigma \sim 0.37$ for all our data.
This further indicates a MM--$\epsilon\chi$PT equivalence
for $SU(2)$ at $\mu\neq0$.

We thank P. Damgaard, M.-P. Lombardo, H. Markum, J.C. Osborn, 
K. Splittorff, J.J.M. Verbaarschot and T. Wettig 
for useful conversation. 
This work was supported by BRIEF award 707, EU network ENRAGE
MRTN-CT-2004-005616, EPSRC grant 
EP/D031613/1 (G.A.), and DFG grant JA483/22-1 (E.B.).

\end{document}